\documentclass[letter,twocolumn]{jpsj3}
\usepackage{txfonts}
\usepackage{tabularx}
\usepackage{graphicx}
\usepackage{dcolumn}
\usepackage{bm, color}
\usepackage{amsmath}
\usepackage{amsfonts}
\usepackage{amssymb}
\usepackage{multirow}
\providecommand{\U}[1]{\protect\rule{.1in}{.1in}}

\newcommand{\vv}[1]{\boldsymbol #1}
\newcommand{\ket}[1]{\vert{#1}\hspace{0.45pt}\rangle}

\usepackage{physics}

\title{Magnetic Response of Majorana Kramers Pairs Protected by $\mathbb Z_2$ Invariants}

\author{Yuki Yamazaki$^1$
, Shingo Kobayashi$^{2,3,4}$, and Ai Yamakage$^1$}
\inst{$^1$Department of Physics, Nagoya University, Nagoya 464-8602, Japan \\
$^2$Institute for Advanced Research, Nagoya University, Nagoya 464-8601, Japan \\
$^3$Department of Applied Physics, Nagoya University, Nagoya 464-8603, Japan \\
$^4$RIKEN Center for Emergent Matter Science, Wako, Saitama 351-0198, Japan } 

\date{\today}

\abst{
On the surface of time-reversal-invariant topological superconductors, Kramers pairs of Majorana fermions with chiral and crystalline symmetries exhibit completely uniaxial or octupole anisotropic magnetic response.  
This paper reports possible types of magnetic responses of Majorana Kramers pairs with one-dimensional $\mathbb{Z}_2$ invariants defined by crystalline symmetry. 
In particular, the general theory predicts a new type of magnetic response where two Majorana Kramers pairs associated with the $\mathbb Z_2$ invariant show biaxially (quadrupolar) anisotropic magnetic response, which is a novel type of response that is rarely observed in conventional and Majorana fermions. 
}

\begin{document}
\maketitle

\makeatletter
\def\ext@table{}
\makeatother
\makeatletter
\def\ext@figure{}
\makeatother

The Majorana fermion is a long-sought particle that is its own antiparticle.
Some unconventional superconductors with topological numbers called topological superconductors (TSCs) host Majorana fermions on their surfaces as gapless Andreev bound states \cite{Hu1526, Kashiwaya1641, Hasan3045, Qi1057, Tanaka011013, Sato076501, Haim2019a}. 
The emergent Majorana fermion on the surface follows non-Abelian statistics and is completely stable as long as the superconducting gap remains in the bulk.
These novel properties allow us to apply TSCs to fault-tolerant topological quantum computation with Majorana fermions\cite{Nayak1083}.

Many classes of 3D TSCs were discovered based on the concept of symmetry. 
Time-reversal symmetry (TRS) protects degenerate gapless states that form a Kramers pair, which is called a Majorana Kramers pair, on the surface of 3D time-reversal-invariant TSCs; examples include superconducting states in doped topological insulators\cite{Hor057001,Fu097001,Sasaki217001,Sasaki217004,Hashimoto174527,Fu100509,Matano852,Yonezawa123} and Dirac semimetals\cite{Aggarwal3237,Wang3842,Kobayashi187001,Hashimoto014510,Oudah13617,Kawakami041026}. 
In particular, crystalline symmetries define a new type of topological crystalline superconductor (TCSC) \cite{ueno13, Taylor047006, zhang-kane-mele}. 
One-dimensional time-reversal-invariant superconductors (class DIII) without crystalline symmetry are classified by the $\mathbb Z_2$ topological invariant \cite{schnyder, kitaev, ryu}. 
In addition to the above $\mathbb Z_2$ phase, topological phases in the presence of crystalline symmetry have been thoroughly explored for reflection \cite{chiu-yao-ryu, morimoto-furusaki}, all order-2 \cite{shiozaki14}, nonsymmorphic \cite{Shiozaki195413}, and rotational \cite{Benalcazar224503, Fang01944} symmetries.
To study the fundamental nature and possible applications of these 3D TCSCs, 
it is necessary to understand how Majorana fermions on 3D TCSCs respond to an external field. 
For instance, a Majorana Kramers pair exhibits a completely anisotropic (Ising) magnetic response, which is distinct from the response of conventional (complex) spin-1/2 fermions \cite{Sato094504, Chung235301, Nagato123603, shindou10,  Mizushima12, Tsutsumi113707, Mizushima022001}, owing to TRS reinforced by crystalline symmetries, i.e., magnetic symmetry \cite{shiozaki14}.

Previously\cite{xiong17, kobayashi}, we revealed the relation among the magnetic-dipole and magnetic-octupole responses of a Majorana Kramers pair; magnetic winding number $\mathbb Z$, which coincides with the number of Majorana fermions; and irreducible representation of the superconducting pair potential. 
By applying this result, one can easily determine which multipole magnetic response occurs in a 3D TCSC associated with the $\mathbb Z$ invariant. 
This method, however, is incomplete: it does not include a Majorana Kramers pair associated with $\mathbb Z_2$ topological invariants protected by crystalline symmetry. 
Recent studies have reported results useful for determining the topological invariants of 3D TCSCs from the symmetry indicators in the normal state \cite{Ono115150, Skurativska, Ono013012, Ono09634, Geier11271}. 
In the present paper, we extend these studies by systematically elucidating the magnetic response of Majorana Kramers pairs with one-dimensional invariants $\nu[U] \in \mathbb{Z}_2$ protected by an order-2 symmorphic or nonsymmorphic symmetry $U$. 
In the symmorphic case, 
an energy gap occurs in the Majorana Kramers pair when an applied magnetic field breaks the symmetry $U$. 
Surprisingly,
two Majorana Kramers pairs protected by nonsymmorphic symmetry show a highly anisotropic magnetic response, which is expressed by a quadrupolar-shaped energy gap depending on the magnetic field. 

\textbf{Effective theory for Majorana Kramers pairs on a surface of 3D TSCs.}
We discuss the magnetic responses of Majorana Kramers pairs by an effective theory for them on a surface of 3D TSCs. 
In general, $\mathbb Z_2$ topological invariants correspond to the parity of Majorana Kramers pairs thus there cannot exist two or more pairs stably. However, if the system has a glide plane, there possibly exist two Majorana Kramers pairs\cite{Shiozaki195413, Daido227001}.
Therefore, we firstly reveal the dependence of the magnetic responses of Majorana Kramers pairs on the number of pairs.
$N$ Majorana Kramers pairs are described by Majorana operators $\gamma_1, \cdots, \gamma_{2N}$ satisfying $
\gamma^\dag_i = \gamma_i$ and $
\{\gamma_i, \gamma_j\} = 2 \delta_{ij}
$.
$\gamma_{2n-1}$ and $\gamma_{2n}$ form a Majorana Kramers pair on a surface of 3D TSCs.
The time reversals are given by $\gamma_{2n-1} \to \gamma_{2n}$ and $\gamma_{2n} \to -\gamma_{2n-1}$.
Then, we obtain the coupling between an external field and $N$ Majorana Kramers pairs on the surface by antisymmetric matrices as
\begin{align}
J = \frac{i}{2}\bm\gamma^{\mathrm T} A \bm\gamma,
\
A^{\mathrm T} = - A,
\
A^* = -A.
\end{align}

A single Majorana Kramers pair $\bm\gamma = (\gamma_1, \gamma_2)^{\mathrm T}$ hosts only one operator
$A = s_y$,
where $s_i \ (i=0,x,y,z)$ denotes the $i$th Pauli matrix.
The operator $J$ is time-reversal-odd (magnetic) and coupled to a magnetic field $\boldsymbol{B}$.
Thus, the Hamiltonian for a single Majorana Kramers pair under a magnetic field $\boldsymbol{B}$ has the form $H_{\mathrm{MF}} = f(\boldsymbol{B}) s_y$ 
and the gapped energy spectrum $E_{\text{M}} = f(\boldsymbol{B})$, where
$f(\boldsymbol{B})$ is an analytic odd function:\cite{Dumitrescu245438}
\begin{align}
f(\vv{B}) = \sum_{i} \rho_i B_i + \sum_{i,j,k}\rho_{ijk} B_iB_jB_k + \order{B^5}.
\label{f}
\end{align}
Here, the coefficients $\rho_i$ and $\rho_{ijk}$ depend on the system parameters. 

Two Majorana Kramers pairs, on the other hand, can form four magnetic ($A_1$, $A_2$, $A_3$, and $A_4$) and two electric ($B_1$ and $B_2$) operators. 
They are represented by $A_1 = s_y \tau_0$, $A_2 = s_y \tau_z$, $A_3 = s_0 \tau_y$, $A_4 = s_y \tau_x$, $B_1 = s_z \tau_y$, and $B_2 = s_x \tau_y$ in the basis of $(\gamma_1, \gamma_2, \gamma_3, \gamma_4)$, where $s_i$ acts within a Majorana Kramers pair so that time reversal is given by $A \to \Theta A \Theta^{-1}$ with $\Theta = is_y \mathcal K$.
The effective Hamiltonian is given by
$H_{\text{MF}} = \sum_{i=1}^{4} A_i f_i(\vv B) + \sum_{i=1}^{2} B_i g_i(\vv{B})$,
where $f_i(\vv B)$ is an analytic odd function that has the same form as Eq. (\ref{f}) and $g_i(\vv B)$ is an analytic even function of magnetic fields. 
By diagonalizing the above matrix, the energy gap is obtained as
\begin{align}
\nonumber
E_{\text{M}} &= \sqrt{f_1^2+f_3^2+f_4^2} - \sqrt{f_2^2 + g_1^2 + g_2^2}
\\ &
\sim \sqrt{\sum_{ij} \rho_{ij} B_i B_j} - \sqrt{\sum_{ij} \rho'_{ij} B_i B_j}.
\label{E_{MF}4}
\end{align}
Consequently, two Majorana Kramers pairs vanish on applying a magnetic field along any direction. 
However, the induced gap shows a highly anisotropic magnetic response, as demonstrated later. 

Magnetic responses of Majorana Kramers pairs on a surface of 3D TCSCs are constrained by crystalline symmetry in addition to the number of Majorana Kramers pairs discussed above. In the following, we construct a general theory for crystalline $\mathbb{Z}_2$ topological phase and discuss the effects of crystalline symmetry on magnetic responses. Then, we show that various anisotropic magnetic responses such as dipole and quadrupole are realized thanks to crystalline symmetry.

\textbf{Crystalline $\mathbb{Z}_2$ topological phase and magnetic response.} 
In this paper, we focus on Majorana Kramers pairs on a surface of 3D TSCs. Note that, as will be shown later, Majorana Kramers pairs protected by crystalline symmetry are characterized by a topological number defined in a one-dimensional subspace. Hence, the following discussions can be applied to two-dimensional and one-dimensional systems and nodal superconductors. The crystalline symmetry protecting Majorana Kramers pairs which appear on a surface of 3D TCSCs is defined below.
First of all, we introduce the Hamiltonian and symmetry operation, 
following which we classify the possible magnetic responses of Majorana Kramers pairs on 3D TCSCs. 

The BdG Hamiltonian for time-reversal-invariant three-dimensional superconductors has the form
\begin{align}
 H(\boldsymbol k) &= \mqty(
  h(\boldsymbol k) - \mu & \Delta(\boldsymbol k)
  \\
  \Delta(\boldsymbol k) & -h(\boldsymbol k) + \mu
 )
 = \qty[h(\boldsymbol k)-\mu] \tau_z + \Delta(\boldsymbol k) \tau_x,
\end{align}
in the basis of $(c_{\boldsymbol k \uparrow}, c_{\boldsymbol k \downarrow}, c_{-\boldsymbol k \downarrow}^\dag, -c_{-\boldsymbol k \uparrow}^\dag)$, where $\uparrow$ and $\downarrow$ denote the up and down spins, respectively, and the indices for the orbital and sublattice degrees of freedom are implicit. 
The Hamiltonian satisfies TRS, which is expressed as $\Theta H(\vv{k})\Theta^{-1} = H(-\vv{k})$; particle--hole $C$ symmetry (PHS), expressed as $C H(\vv{k}) C^{-1} = -H(-\vv{k}), \ C = \tau_y \Theta$; and chiral $\Gamma$ symmetry, expressed as $\{\Gamma,H(\vv{k})\}= 0, \ \Gamma = \Theta C = \tau_y$.
When the system is invariant against a symmetry operation $g = \{R_g | \boldsymbol{\tau}_g \}$ of a space group, 
which consists of a rotation/screw axis or reflection/glide plane $R_g$ followed by the translation $\boldsymbol{\tau}_g$, 
the Hamiltonian $h(\boldsymbol{k})$ in the normal state satisfies $D^{\dagger}_{\vv{k}}(g) h(\vv{k}) D_{\vv{k}}(g) = h(g\vv{k})$, where $D_{\boldsymbol{k}}(g)$ is the representation matrix of $g$ 
and the momentum $\boldsymbol{k}$ is transformed to $g\vv{k}$. 
The pair potential $\Delta(\vv{k})$, on the other hand, satisfies $D^{\dagger}_{\vv{k}}(g) \Delta(\vv{k}) D_{\vv{k}}(g) = \chi (g) \Delta(g\vv{k})$,
where $\chi(g)$ is the character of $g$ for the one-dimensional representation of the pair potential \cite{multi}. 
Then, the BdG Hamiltonian is invariant, $\tilde{D}^{\dagger}_{\vv{k}}(g) H(\vv{k}) \tilde{D}_{\vv{k}}(g) = H(g\vv{k})$, for $\tilde{D}_{\vv{k}}(g) = \text{diag}[D_{\vv{k}}(g),\chi(g)D_{\vv{k}}(g)]$. 
The particle--hole and chiral transformations of the representation matrices depend on the character:
\begin{align}
\tilde{D}^{\dagger}_{-\vv{k}}(g) C \tilde{D}_{\vv{k}}(g) &= \chi(g)C,
\label{particle-hole} \\
\tilde{D}^{\dagger}_{\vv{k}}(g) \Gamma \tilde{D}_{\vv{k}}(g) &= \chi(g)\Gamma,
\label{chiral}
\end{align}
where we use $\Theta \tilde{D}_{\bm k}(g) \Theta^{-1} = \tilde{D}_{-\bm k}(g)$.

The square of the representation matrix is given by
\begin{align}
 D^2_{\boldsymbol{k}}(g) = -e^{-i \boldsymbol{k} \cdot (R_g \boldsymbol{\tau}_g + \boldsymbol{\tau}_g)}
\end{align}
and classified into $D^2_{\boldsymbol{k}}(g)=-1$ and 1 for time-reversal-invariant momenta (TRIMs). 
$D^2_{\boldsymbol{k}}(g)=-1$ holds when $g$ represents twofold rotation and reflection. 
The twofold screw axis and glide plane also satisfy $D^2_{\boldsymbol{k}}(g)=-1$ for $\boldsymbol{k} \cdot 2 \boldsymbol{\tau}_g = 0$.
On the one hand, $D^2_{\boldsymbol{k}}(g)=1$ holds when $g$ represents the screw axis and glide plane on the zone boundary with $\bm k \cdot 2\tau_{g} = \pi$ because $\boldsymbol{\tau}_g$ can be a half translation vector. 
We call $D^2_{\boldsymbol{k}}(g)=-1$ and 1 symmorphic and nonsymmorphic symmetry, respectively. 

Symmetry that preserves the surface, which is referred to as surface symmetry, can protect Majorana Kramers pairs. 
To formulate the surface symmetry, 
we denote the momentum by $(k_\perp, \boldsymbol{k}_\parallel)$, which are components perpendicular and parallel to the surface, respectively.
Next, $\boldsymbol{k}_\parallel$ is fixed to a TRIM, where Majorana Kramers pairs can appear with zero energy, and omitted. 
The surface symmetry $U$ satisfies 
\begin{align}
 [D_{k_\perp}(U), h(k_\perp)] = 0.
\end{align}
The symmetry can divide the BdG Hamiltonian $H$ into parts in the eigenspaces of $\tilde{D}_{k_\perp}(U)$ as $H = H_+ \oplus H_-$, where the subscript $\pm$ denotes the eigenvalue of $\tilde D_{k_\perp}(U)$. 
Majorana fermions in $H_+$ and $H_-$ do not hybridize with each other. 
In other words, they are protected by the symmetry $U$. 
Note that the representation matrix $D_{k_\perp}(U)$ must be independent of $k_\perp$ in order to protect Majorana Kramers pairs, i.e., $U$ is the reflection perpendicular, glide plane consisting of the reflection perpendicular and translation parallel, or twofold rotation perpendicular to the surface.
Hereafter, the subscript $k_\perp$ is omitted.

Next, we derive a condition on the representation matrix and character to protect the $\mathbb Z_2$ topological phase with Majorana Kramers pairs and relate it to the magnetic responses. 
According to Eq. (\ref{particle-hole}), $H_\pm$ has particle--hole symmetry for $\chi(U) = D^2(U) = \pm 1$ because the eigenvalues of $\tilde D(U)$ are $\pm 1$ and $\pm i$ for $D^2(U) = 1$ and $-1$, respectively. 
The magnetic responses are classified into three types, (A)--(C), as summarized in Table \ref{fivetype}, which is the main result of this paper.

\begin{table*}[t]
	\centering
	\caption{
		Three types, (A)--(C), of magnetic responses of Majorana Kramers pairs. 
		The square of the representation matrix $D^2(U)$ and character $\chi(U)$ of symmetry operation $U$, which protects Majorana Kramers pairs; 
		symmetry class (Class) and one-dimensional topological invariant (Topo);
		number of Majorana Kramers pairs (\#MKP);
		and
		the energy spectrum $E_{\mathrm M}(\boldsymbol{B})$ under a magnetic field $\boldsymbol{B}$ are shown. 
		$\boldsymbol{n}$ denotes the direction perpendicular to the mirror plane when $U$ is a reflection/glide plane or the direction parallel to the rotational axis when $U$ is a twofold rotation.
	} 
	\begin{tabular}{cccccccc}
		\hline\hline
		Type & $\boldsymbol{k}_\parallel \cdot 2\boldsymbol{\tau}_g$ & $D^2(U)$ & $\chi(U)$ & Class & Topo & \#MKP & $E_{\mathrm M}(\boldsymbol{B})$ 
		\\
		\hline
		A & 0 & $-1$ & $-1$ & D & $\mathbb Z_2$ & 1 & $\sum_i\rho_i B_i +  \sum_{ijk}\rho_{ijk} B_iB_jB_k$ with $E_{\mathrm M}(B \boldsymbol{n}) = 0$
		\\
		B & $\pi$ & 1 & 1 & DIII & $\mathbb Z_2 \oplus \mathbb Z_2$ & 1 & $\propto \boldsymbol n \cdot \boldsymbol{B}$ 
		\\
		C & $\pi$ & 1 & 1 & DIII & $\mathbb Z_2 \oplus \mathbb Z_2$ & 2 & $\sqrt{ \sum_{ij} \rho_{ij} B_i B_j} - \sqrt{ \sum_{ij} \rho'_{ij} B_i B_j}$
		\\
		\hline\hline
	\end{tabular}
	\label{fivetype}
\end{table*}

Type (A). 
For $\chi(U)=-1$ and $D^2(U)=-1$, $H_\pm$ breaks TRS (class D).
A single Majorana Kramers pair in $H=H_+ \oplus H_-$ is divided into Majorana fermions in $H_+$ and $H_-$, which are associated with the $\mathbb Z_2$ invariant:
\begin{align}
 \nu_\pm[U] &= \int_{-\pi}^\pi \frac{dk_\perp}{\pi} a_\pm(k_\perp) \mod 2, 
 \\
 a_\pm(k_\perp) &= -i\sum_{n \in \text{occ}} \mel{k_\perp n \pm}{\partial_{k_\perp}}{k_\perp n \pm},
\end{align} 
where the summation is taken over all occupied states and $H_\pm(k_\perp) \ket{k_\perp n \pm} = E_{n\pm}(k_\perp)\ket{k_\perp n \pm}$.  
Because $H_+$ and $H_-$ are switched by time reversal, $\nu[U] := \nu_+[U] = \nu_-[U]$ holds. 
A magnetic field along 
$\bm n$ is perpendicular to the mirror when $U$ is a reflection/glide plane or is parallel to the rotational axis when $U$ is a twofold rotation because the applied magnetic field $B \bm n$ keeps the symmetry $U$. 
Then, the $\mathbb Z_2$-invariant $\nu[U]$ remains well-defined under the magnetic field $B \boldsymbol{n}$ and has the same value as that under zero field.  
Therefore, we find that the energy gap $E_\text{M}$ is expressed by Eq. (\ref{f}) as $E_\text{M} = f(\bm B)$, satisfying $f(B\bm n)=0$.

For $\chi(U)=1$ and $D^2(U)=1$, on the other hand, $H_\pm$ respects TRS (class DIII) and can have a single Majorana Kramers pair, which is associated with
\begin{align}
 \nu_\pm[U] = \int_{-\pi}^{\pi} \frac{dk_\perp}{2\pi} a_{\pm}(k_\perp) \mod 2,
\end{align} 
with the gauge fixed to $\Theta \ket{k_\perp 2n-1 \pm } = \ket{-k_\perp 2n \pm }$. 
The whole system can be classified into types (B) and (C) as follows.

Type (B).
This type corresponds to a single Majorana Kramers pair with $(\nu_+[U], \nu_-[U])=(1,0)$ or $(0,1)$, which is protected by the nonsymmorphic symmetry $U$ for $\chi(U)=D^2(U)=1$. 
In reality, it is protected by the magnetic glide-plane symmetry $\Theta[U] = \tilde{D}(U) \Theta$ with $\Theta^2[U] = -1$, which is regarded as TRS in the whole system with $H=H_+ \oplus H_-$. 
An energy gap occurs when the direction $\bm n$ is perpendicular to the glide plane because the applied magnetic field $B \bm n$ breaks the magnetic glide-plane symmetry.
The resulting energy gap is proportional to $E_\text{M} \propto \bm n \cdot \bm B$.

Type (C). 
This type corresponds to two Majorana Kramers pairs with $(\nu_+[U], \nu_-[U])=(1,1)$, which is protected by the nonsymmorphic symmetry $U$ for $\chi(U)=D^2(U)=1$. 
The energy gap of the Majorana Kramers pairs is given by Eq. (\ref{E_{MF}4}).

\begin{figure}
	\centering
	\includegraphics[scale=0.31]{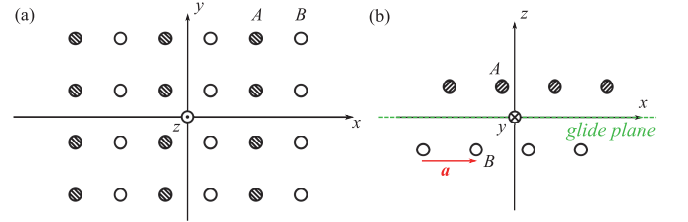}
	\caption{(a) Top and (b) side views of the lattice structure of the toy model. 
		$\boldsymbol{a}$ represents the primitive translational vector along the $x$ axis. 
		There exist two sublattices denoted by A (closed circles) and B (open circles). 
		The $(xy)$ plane (the dashed green line) is the glide plane.}
	\label{model}
\end{figure}

\textbf{Model Hamiltonian and numerical results.}
We finally verify the type (C) by examining a toy model on a layered 2D lattice which has two sublattices with the space group $Pmma$ (No. 51) including a glide plane $U$ on the $(xz)$ surface\cite{wang-liu}, as shown in Fig. \ref{model}. 
The $B_{3u}$ pairing, which is equivalent to $p_x$-wave pairing, satisfies $\chi(U) =D^2(U) = +1$ at $k_x=\pi$ and then the magnetic response possibly is of type (C). Hence, we focus on the $B_{3u}$ pairing for $k_x = \pi$. 

\begin{figure}
\centering
\includegraphics[scale=0.30]{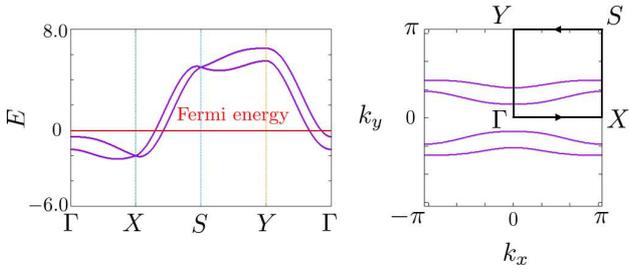}
\caption{Energy dispersion (left) and Fermi surface (right) of the normal state $h(\vv k)$. }
\label{band}
\end{figure}

Here, we set the lattice constants as $1$. 
The normal part $h(\vv k)$ of the BdG Hamiltonian is 
\begin{align}
\nonumber
h(\vv k) &= c(\vv k)\sigma_0 s_0 + t_3 \cos(k_x/2) \sigma_1(k_x) s_0 
\\ & \quad + (\lambda_1 s_x \sin k_y + \lambda_2  s_y \sin k_x) \sigma_3,
\\
c(\vv k) &= m_0 + t_1 \cos k_x + t_2 \cos k_y,
\end{align}
and the pair potential $\Delta(\bm k)$ for the $B_{3u}$ pairing is
\begin{align}
\Delta(\bm k) = \Delta \sigma_0 s_z \sin k_y 
+ \Delta' \sigma_2(k_x) s_x \sin(k_x/2) \sin k_y,
\end{align}
where $s$ and $\sigma$ denote the Pauli matrices representing the spin and layer degrees of freedom (A and B), respectively. 
Here, we introduce the modified Pauli matrices
\begin{align}
\sigma_1(k_x) = 
\pmqty{
0 & e^{ik_x/2} \\
e^{-ik_x/2} & 0
},
\
\sigma_2(k_x) 
=
\pmqty{
0 & -ie^{ik_x/2} \\
ie^{-ik_x/2} & 0
},
\end{align}
and $\sigma_3=\text{diag}(1,-1)$.
When the pair potential is smaller than spin-orbit coupling, i.e., $\lambda^2_1 > \Delta^2 + {\Delta'}^2 $, nodes exist at $k_x=\pi$. 
Hereafter, we assume the gapped case $\lambda^2_1 < \Delta^2 + {\Delta'}^2 $, where Majorana Kramers pairs appear.  
The band structure of the normal state $h(\vv k)$ is shown in Fig. \ref{band}. 
The parameters are set to those listed in Table \ref{parameter}.
All the bands are twofold degenerate at any momentum owing to inversion and time-reversal symmetries. 
Particularly, the bands are fourfold degenerate at the $X$ and $S$ points because of those symmetries and the glide-plane symmetry.  

\begin{table}
\centering
\caption{Parameters of the numerical model.}
\begin{tabular}{cccccccccc}
\hline\hline
$m_0$ & $t_1$ & $t_2$ & $t_3$ & $\lambda_1$ & $\lambda_2$ & $\Delta$ & $\Delta'$ & $|\vv{B}|$  \\ 
\hline 
$2.0$ & $0.5$ & $-3.5$ & $0.5$ & $0.7$ & $-0.5$ & $1.2$ & $0.6$ & $0.1$  \\
\hline\hline
\end{tabular}
\label{parameter}
\vspace{-2.5mm}
\end{table}

The Hamiltonian with the $(xz)$ surfaces has the form 
\begin{align}
\nonumber
 H(k_x) &= \sum^{N_y}_{n=1}c_n^{\dagger}(k_x)
  \epsilon(k_x)
 c_n(k_x)
 \nonumber\\ & \quad 
+ \sum^{N_y-1}_{n=1} 
\qty[
c_n^{\dagger}(k_x)
 t_y(k_x) 
	c_{n+1}(k_x)
 + \mathrm{h.c.}
 ], 
 \label{tight}
\end{align}
where $N_y$ denotes the number of sites along the $y$ direction. 
The Fermi energy is set to $0$. The system has four Fermi surfaces between the $X$ and $S$ points. 
The onsite energy $\epsilon(k_x)$ and hopping $t_y(k_x)$ are defined by
$
 \epsilon(k_x)
 = (m_0 + t_1\cos k_x) \sigma_0 s_0 \tau_z
 + t_3 \cos(k_x/2) \sigma_1(k_x) s_0 \tau_z
  + \lambda_2 \sin k_x \sigma_3 s_y \tau_z
$ and
$ t_y(k_x)
 = 
 (t_2/2)\sigma_0 s_0 \tau_z - i(\lambda_1/2) \sigma_3 s_x \tau_z 
  -i(\Delta/2)\sigma_0 s_z \tau_x - i(\Delta'/2) \sin(k_x/2) \sigma_2(k_x) s_x \tau_x$, respectively.

\begin{figure}
\centering
\includegraphics[scale=0.23]{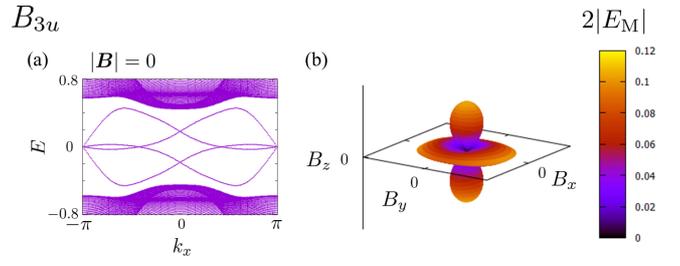}
\caption{(a) Energy spectrum of Eq. (\ref{tight}) without a magnetic field.
	Two Majorana Kramers pairs are located at $k_x=\pi$. 
	They are gapped by an external magnetic field. 
	(b) The polar plot of the energy gap $|E_{\mathrm M}|$ of $H+H_{\mathrm{Z}}$ as a function of $\boldsymbol{B}$ with $|\vv{B}|=0.1$.}
\label{mag}
\vspace{-2.5mm}
\end{figure}

\begin{figure}
\centering
\includegraphics[scale=0.23]{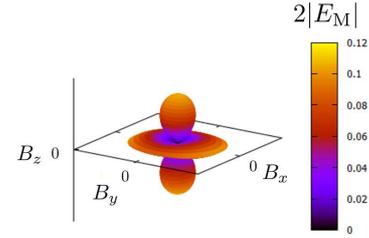}
\caption{The polar plot of the function Eq. (\ref{fitting}) with $|\vv{B}|=0.1$. The shape is almost the same as Fig. \ref{mag}(b).}
\label{mag2}
\vspace{-3.5mm}
\end{figure}

A magnetic field induces the Zeeman term
\begin{align}
H_{\text{Z}}(k_x) = \sum^{N_y}_{n=1} c_n^{\dagger}(k_x) \vv{B} \cdot \vv{s} \sigma_0 \tau_0 c_n(k_x).
\end{align}
The energy spectrum of $H(k_x) + H_{\text{Z}}(k_x)$ is shown in Fig. \ref{mag}. 
Without a magnetic field, i.e., with $|\vv B|=0$ [Fig. \ref{mag}(a)], the superconducting gap in the bulk is of the order of $\sim 1.4$. 
At $k_x = \pi$, two Majorana Kramers pairs exist with zero energy on the $(xz)$ surface. 
Figure \ref{mag}(b) shows a polar plot of the energy gap $E_{\mathrm{M}}(\boldsymbol{B})$ with $|\vv{B}|=0.1$. 
Magnetic fields along any direction destroy the two Majorana Kramers pairs and yield a gap of the order of $\sim 0.12$ at maximum. 
Here, we confirm that Fig. \ref{mag}(b), which is the magnetic response of type (C), can be reproduced by Eq. (\ref{E_{MF}4});
\begin{align}
E_{\text{M}}(\boldsymbol{B}) 
=\sqrt{\rho_{xx}B^2_x + \rho_{yy}B^2_y + \rho_{zz}B^2_z} - \sqrt{\rho'_{zz} B^2_z},
\label{fitting}
\end{align}
where the coefficients are  $\rho_{xx} = 0.197, \ \rho_{yy} = 0.268, \ \rho_{zz} = 0.10, \ \rho'_{zz} = 0.709$ and all other coefficients are $0$.
Figure \ref{mag2} shows a polar plot of the function Eq. (18) with $|\vv{B}|=0.1$. 
By comparing Figs. \ref{mag}(b) and \ref{mag2}, one can find that they have almost the same shape. 
In conclusion, we verify that the magnetic response of type (C) appear on the $(xz)$ surface of a layered 2D lattice with the space group $Pmma$ when the pair potential is $B_{3u}$ pairing, i.e., the energy gap $E_{\mathrm{M}}(\boldsymbol{B})$ mimics the anisotropy of a quadrupole which is allowed within Eq. (3).
This behavior is entirely different from those of other Majorana and complex fermions. 

\textbf{Discussion}. 
We found three types of the magnetic responses in Majorana Kramers pairs with the $\mathbb{Z}_2$ invariants, which depend on the surface symmetry and the number of Majorana Kramers pairs.
In type (A), i.e., the symmorphic case, there exists only a single Majorana Kramers pair, in which
an external magnetic field creates a uniaxially anisotropic gap. 
On the other hand, in types (B) and (C), the case of nonsymmorphic symmetry,  
the magnetic response depends on the number of Majorana Kramers pairs:
(B) a single Majorana Kramers pair behaves as an Ising spin, which is the same as the response associated with the $\mathbb Z$ invariant. 
(C) two Majorana Kramers pairs show a biaxially (quadrupolar) anisotropic magnetic response, which is a novel type of response rarely observed in conventional and Majorana fermions.

This prediction was verified in a bilayer model with a glide plane ($Pmma$). 
In fact, our result can be applied to materials with any nonsymmorphic space group,\cite{paper2} such as 
UCoGe, 
which has the crystalline symmetry of the space group $Pnma$\cite{CANEPA1996225}. 
Previous studies have strongly suggested that this material is a ferromagnetic superconductor at ambient pressure \cite{Huy067006, Aoki061011} and a time-reversal-invariant one at high pressure \cite{Hassinger073703, Slooten097003, Bastien125110, Manago020506,Cheung134516,Mineev104501}. 
A theoretical study \cite{Daido227001} reported that, based on an experimental result\cite{Hattori066403}, the ferromagnetic superconducting state has $A_u$ symmetry of $C_{2h}$ 
and the state can deform into either $A_u$ or $B_{1u}$ symmetry of $D_{2h}$ at high pressure. 
The $B_{1u}$-pairing (glide-even $\chi(G_n)=1$) state hosts two Majorana Kramers pairs on the
$(0\bar{1}1)$ surface, while the $A_u$-pairing (glide-odd $\chi(G_n)=-1$) hosts no Majorana Kramers pair 
because the surface has only $G_n$ symmetry satisfying $D_{(k_x, \pi, \pi)}^2(G_n)= \chi(G_n) = 1$ for the $B_{1u}$ pairing. 
Thus, the magnetic response of Majorana Kramers pairs on UCoGe is predicted to be of type (C). 
%
The gap induced in the Majorana Kramers pairs may be observed through surface tunneling spectroscopy under a magnetic field or with a magnet attached \cite{tanaka95, fogelstrom97, tanaka02, tanuma02, tanaka09, tamura17}.

$C_3$, $C_4$, and $C_6$ symmetries \cite{Fang01944}, which are beyond the scope of this paper, might realize new types of magnetic responses. 
The discussions in this paper suggest that multiple Majorana Kramers pairs can be active against electric perturbations. Therefore, we also need to clarify the $\mathbb{Z}$ and $\mathbb{Z}_2$ topological invariants and the electric responses of multiple Majorana Kramers pairs. 
These issues will be addressed in a future paper.

\begin{acknowledgments}
This work was supported by Grants-in-Aid for Scientific Research on Innovative Areas ``Topological Material Science" (Grant No. JP18H04224) from JSPS of Japan. S. K. was supported by JSPS KAKENHI Grants No. JP19K14612 and by the CREST project (JPMJCR16F2) from Japan Science and Technology Agency (JST).
\end{acknowledgments}

\bibliographystyle{jpsj}
\bibliography{ref}

\begin{thebibliography}{10}
\bibitem{multi}
A multidimensional irreducible representation is regarded as one-dimensional
  irreducible representations of the subgroup.
\end{thebibliography}

\clearpage

\end{document}